%% file: zram-hotnets.tex
\documentclass[sigconf, 10pt, nonacm]{acmart}
\usepackage{hyperref}
\usepackage{tabularx}
\renewcommand{\arraystretch}{1.1}

\hypersetup{pdfstartview=FitH,pdfpagelayout=SinglePage}

\usepackage{kim-paper-macros}

\usepackage{amsmath}
\usepackage{bm}
\usepackage{tikz}
\usepackage{pgfplots}
\pgfplotsset{compat=1.16}

\usepackage[capitalise,nameinlink]{cleveref}
\crefname{algorithm}{Algorithm}{Algorithms}
\Crefname{algorithm}{Algorithm}{Algorithms}
\crefname{section}{\S\!}{\S\S\!}
\Crefname{section}{Section}{Sections}
\crefname{figure}{Figure}{Figures}
\Crefname{figure}{Figure}{Figures}
\crefname{equation}{Equation}{Equations}
\Crefname{equation}{Equation}{Equations}
\crefname{listing}{Listing}{Listings}
\Crefname{listing}{Listing}{Listings}
\crefname{defn}{definition}{definitions}

\renewcommand{\sectionautorefname}{\S\kern-0.2em}
\renewcommand{\subsectionautorefname}{\S\kern-0.2em}
\renewcommand{\subsubsectionautorefname}{\S\kern-0.2em}

\usepackage[inline]{enumitem}
\newlist{inlist}{enumerate*}{1}
\setlist[inlist]{label=\textbf{(\arabic*)}}

\newcommand{\cram}{\textsc{cram}\xspace}
\newcommand{\zram}{\textsc{zram}\xspace}
\newcommand{\kap}{\kappa}
\newcommand{\rmax}{\rho_{\max}}

\newcommand{\figplaceholder}[3][\columnwidth]{%
  \IfFileExists{#2.pdf}{\includegraphics[width=#1]{#2}}{%
    \IfFileExists{#2.png}{\includegraphics[width=#1]{#2}}{%
      \setlength{\fboxsep}{6pt}%
      \fbox{\parbox[c][4.8cm][c]{0.86#1}{\centering\footnotesize\itshape #3}}}}}

\begin{document}

\pagenumbering{arabic}

\title{Towards Simple Models of Complex SmartNICs}

\author{Robert Chang}
\affiliation{%
  \institution{University of California, Los Angeles}
  \country{}
}

\author{Teng Jiang}
\affiliation{%
  \institution{University of California, Los Angeles}
  \country{}
}

\author{Wonsup Yoon}
\affiliation{%
  \institution{The University of Texas at Austin}
  \country{}
}

\author{Daehyeok Kim}
\affiliation{%
  \institution{The University of Texas at Austin}
  \country{}
}

\author{Sam Kumar}
\affiliation{%
  \institution{University of California, Los Angeles}
  \country{}
}

\author{George Varghese}
\affiliation{%
  \institution{University of California, Los Angeles}
  \country{}
}

\input{sections/abstract}

\maketitle
\hypersetup{pdfauthor={}}

\input{sections/introduction}
\input{sections/zram}
\input{sections/platform}
\input{sections/ddos}
\input{sections/cases}
\input{sections/patterns}
\input{sections/related}
\input{sections/open}
\input{sections/conclusion}

\label{EndOfPaper}

{\bibliographystyle{plain}
\bibliography{reference}}

\end{document}

%% file: sections/abstract.tex
\begin{abstract}

Cloud vendors push ambitious in-network processing (e.g., crypto, telemetry) onto the NIC to offload servers even as link rates climb to terabit speeds. Vendors have responded with heterogeneous SmartNICs. For example, NVIDIA BlueField-3 interposes\,---\,between the wire and the host CPUs\,---\,a line-rate eSwitch, a multithreaded Data-Path Accelerator, general-purpose ARM cores, and a sea of fixed-function accelerators. These devices are notoriously hard to program, and harder still to predict. Applications can be implemented in many ways, with each choice potentially hitting a different bottleneck. A designer ideally needs to know\,---\,cheaply, and before a line of code is written\,---\,feasible choices and their bottlenecks, and design patterns to improve performance.

Our paper offers a starting point to answer these questions using what we call the \zram model. It pairs a \emph{platform graph} of processing zones and their channels with a \emph{program graph} of tasks and their traffic fractions. The application designer or a compiler chooses a \emph{placement} that maps the program graph onto the platform graph. Three metrics computed directly from this mapping\,---\,capability, roofline, and capacity\,---\,score the placement, deciding its feasibility and naming the bottleneck resource. We use a DDoS detector as a primary case study, and briefly explore two other applications, decision-tree inference and RDMA traversal. We distill seven design patterns for programming SmartNICs including a key one we call \emph{sifting}. \zram generalizes to other SmartNICs such as Intel IPU E2200 and AMD Pensando Salina 400, and opens a new research agenda that includes compilers and hardware design.
\vspace{-1em}
\end{abstract}

%% file: sections/introduction.tex
\section{Introduction}
\label{sec:introduction}

{\em \ldots build models that capture the essential core of a problem rather than its literal messiness.}\,---\,Paul Samuelson
\smallskip

SmartNICs are now core cloud infrastructure. As link rates climb toward 800 Gbps, cloud vendors offload infrastructure work\,---\,virtual switching, encryption, firewalling, and DDoS protection\,---\,to these programmable cards, reclaiming host cores to rent to tenants and filtering threats before they reach host memory. Programmability lets providers such as AWS Nitro~\cite{nitro} and Microsoft Azure~\cite{accelnet} roll out new protocols and tenant isolation in firmware, without swapping hardware. But doing all this at line rate has made NICs remarkably heterogeneous, complex, and messy.

Modern SmartNICs mix general-purpose CPUs, specialized accelerators, tiers of fast and slow memory, and interconnects. 
\emph{Where} a task runs matters as much as \emph{what} it does. For example, NVIDIA's BlueField-3 (BF-3)~\cite{bluefield_3} packs four kinds of processing between the network port and the server it plugs into: a match-action eSwitch that filters and forwards at 400 Gbps, a 256-thread RISC-V Data-Path Accelerator (DPA) directly in the fast path, 16 general-purpose ARM cores backed by their own DDR5 memory, and an assortment of fixed-function accelerators for encryption, compression, etc. Any nontrivial NIC program can be mapped onto this hardware in many ways, with no simple way to judge the best choice other than time-consuming implementation.

\smallskip\noindent\textbf{Motivating example:}
Consider a hierarchical DDoS detector for distinguishing innocuous traffic spikes from genuine attacks: meter each destination's rate, group the sources of over-rate destinations into buckets to gauge diversity, and count unique sources on the suspicious destinations. \emph{Where should each stage run?} Metering is cheap and touches every packet, so placement on the eSwitch is natural. The unique-source sketch is stateful, but whether its working set fits in the DPA's on-chip cache governs the sustainable packet-processing rate. Single-zone placements are dead ends. All-eSwitch is unworkable, since the restrictive match-action pipeline cannot express per-packet read-modify-write operations that bucketing and diversity counting require. All-ARM or all-host is legal but wasteful, forcing every packet off the fast path for work that is almost always a no-op. The interesting designs spread work across zones, escalating rare, expensive cases off the fast path, but which task lands where is far from obvious.

\smallskip\noindent\textbf{Today's landscape:}
Placement design runs largely on informed guesswork. A designer picks a promising placement, builds it, and measures the result. Whether a design works, and what will limit it, only reveal themselves after the implementation already exists. Previous work targets the implementation side of this problem. Compiler frameworks such as Alkali~\cite{alkali} automatically port and parallelize concrete programs across SmartNICs, freeing the designer from re-implementing for each target. But none answers the core questions \emph{before implementation}: which placements are feasible, and which resources are bottlenecks?

\smallskip\noindent\textbf{Our solution:}
The RAM model~\cite{ram} abstracts a sequential computer, and PRAM~\cite{pram} extends it to many processors sharing memory. Neither is faithful to any real machine, yet both let algorithm designers predict running time within constant factors before implementation. Reasoning about SmartNIC placement needs similar abstraction. We propose \zram, a lightweight \emph{model of computation} for SmartNICs.
The SmartNIC is modeled once as a \emph{platform graph}: processing zones connected by channels. Each zone is annotated with its operation menu, cache threshold, memory bandwidth and capacity, and width. A candidate design is modeled as a \emph{program graph} of tasks and traffic fractions. Simple checks\,---\,capability, roofline, and capacity\,---\,decide feasibility and reveal the bottleneck resource. Like the RAM model, \zram trades precision for cheap, portable prediction: to explore a new design, edit the program graph and recompute; to adapt to hardware changes, edit the platform graph and recompute.

\smallskip\noindent\textbf{Beyond BlueField:}
The placement problem is not unique to NVIDIA. Intel's IPU E2200~\cite{ipu_e2200} combines 24 ARM Neoverse N2 cores with a P4 packet-processing pipeline, and AMD's Pensando Salina 400~\cite{pensando_salina_400} pairs 16 ARM Neoverse N1 cores with 232 P4 Match Processing Units. The architectures differ, but each is a heterogeneous SmartNIC whose many processing zones pose the same placement challenge. \zram applies to all of them by defining a new platform graph for each hardware target. Here, we focus on BF-3 because it has been extensively benchmarked~\cite{demystifying, battle}.

\smallskip\noindent\textbf{A research agenda:}
\zram is a starting point, not a final answer.
Treating placement as a modeling problem opens a line of inquiry that can seed several follow-on efforts: compilers that search the placement space automatically, models that capture cross-tenant contention when many programs share one NIC, cyclic formulations for closed-loop transports such as credit-based flow control, and running \zram in reverse to guide SmartNIC hardware design.

\smallskip\noindent\textbf{Contributions:}
\begin{itemize}[leftmargin=1.1em,topsep=1pt,itemsep=1pt,parsep=0pt]
\item \zram, a model of computation that pairs a platform graph with a program graph and scores a placement with three metrics\,---\,capability, roofline, and capacity\,---\,that decide feasibility and identify the bottleneck. A lightweight simulator computes them from the two graphs.
\item A primary study, a hierarchical DDoS detector, plus two shorter ones: decision-tree inference and RDMA traversal.
\item Seven design patterns for programming heterogeneous SmartNICs, including \emph{sifting}: shrinking the traffic fraction zone by zone, from the wire toward the host.
\end{itemize}

%% file: sections/zram.tex
\section{The \zram Model}
\label{sec:zram}\label{sec:model}

A SmartNIC is a \emph{platform graph} $G_H = (Z, C)$ of zones $Z$ connected by directed channels $C$. A program is a \emph{program graph} $G_P = (V, E)$, a directed acyclic graph (DAG) whose vertices $V$ are tasks and whose edges $E$ carry dataflow. A placement $\mu : V \to Z$ assigns each task to a zone.

\subsection{Platform graph}
Every zone $z \in Z$ is described by five fields.

\textbf{Operation menu} $M_z$: the operations the zone supports, drawn from a shared vocabulary that includes \textsf{match} (exact, LPM, ternary, range), \textsf{meter}, \textsf{counter}, \textsf{hash}, \textsf{encrypt}, \textsf{compress}, \textsf{regex}, and \textsf{general compute}.

\textbf{Cache threshold} $\kap_z$: the working-set size below which the zone runs at its fast cached rate. In practice, a zone's bandwidth falls in steps as its working set grows, because real caches are multi-level. $\kap_z$ marks the step whose crossing incurs an \emph{order-of-magnitude} drop.

\textbf{Memory bandwidth}: the aggregate memory bandwidth the zone can sustain (bits/s). A zone with a modeled $\kap$ carries two\,---\,an \emph{upper} (cached) bound $B_z^{\uparrow}$ and a \emph{lower} (uncached) bound $B_z^{\downarrow}$; every other zone carries a single bound $B_z$.

\textbf{Memory capacity} $\mathit{Cap}_z$: the size of the zone's memory.

\textbf{Width} $w_z$: the number of independent lanes (threads or cores) the zone can run in parallel. The per-lane rate is $B_z/w_z$.

\smallskip
A channel $c \in C$ carries a single field: a channel bandwidth $B_c$ (bits/s). Each channel is unidirectional\,---\,a two-way link counts as two separate channels.
 
\subsection{Program graph}
Every task $v \in V$ is described by four fields.

\textbf{Operation} $\mathrm{op}(v)$: the operation the task
performs, drawn from the same vocabulary as the operation menus.

\textbf{Processing work} $s_v$: the bits the task reads, modifies, and writes to process one packet. For payload-touching tasks $s_v$ is roughly the packet size; for stateful tasks it is the state read and updated per packet, which can far exceed the packet.

\textbf{Memory footprint} $m_v$: the size of the data structures (e.g., lookup tables, bitmaps, counters) the task keeps.

\textbf{Traffic fraction} $\rho_v \in [0,1]$: the share of the line-rate stream that reaches the task. Fractions need not sum to one.

\smallskip
An edge $(u,v) \in E$ carries a single field: a transfer size $t_{u,v}$, the bits moved from $u$ to $v$ per packet. Cross-zone edges consume channel bandwidth; same-zone edges do not.
 
\subsection{Core metrics}\label{sec:metrics}
\noindent\textbf{M1: Capability.} Every task must be placed on a zone whose menu supports its operation.
\begin{equation}
\forall v,\ \mathrm{op}(v)\in M_{\mu(v)}
\label{eq:capability}
\end{equation}
\textsf{General compute} admits any operation, at the software rate.

\smallskip\noindent\textbf{M2: Roofline.} A placement is feasible only if every zone and channel can sustain its offered load. Let $R_{\text{line}}$ be the line rate in packets per second, so the packet rate reaching task $v$ is $\rho_v R_{\text{line}}$ and its memory-traffic demand is $\rho_v R_{\text{line}} s_v$ bits/s. Because $s_v$ counts every bit the task reads, modifies, and writes per packet, this demand can exceed the line \emph{bitrate} whenever a task touches more state than the packet carries.

\emph{Zones.} A zone's sustained rate depends on whether its \emph{working set} $S_z=\sum_{v:\mu(v)=z} m_v$ stays cached:
\begin{equation}
B_z^{\text{eff}}=
\begin{cases}
B_z^{\uparrow} & \text{if } S_z \le \kap_z\quad(\text{cached}),\\[3pt]
B_z^{\downarrow} & \text{if } S_z > \kap_z\quad(\text{uncached}).
\end{cases}
\label{eq:resolve}
\end{equation}
Taking $S_z$ to be the full footprint rather than the actively touched hot set is deliberately conservative.
Since a task's hot set is no larger than its footprint $m_v$, this can only push a zone past $\kap_z$, never below, so \zram never overstates a zone's sustained rate.
Width sets how much of $B_z^{\text{eff}}$ a task can reach.
A data-parallel task spreads its packets across all $w_z$ lanes and draws on the zone's aggregate rate, whereas a serial task\,---\,a dependent per-packet chain\,---\,is pinned to one lane and sees only $B_z^{\text{eff}}/w_z$.
A feasible placement must satisfy the aggregate bound over all co-located tasks and, for each serial task, its per-lane bound:
\begin{align}
\sum_{v:\mu(v)=z}\rho_v\, R_{\text{line}}\, s_v &\;\le\; B_z^{\text{eff}}, \label{eq:roof-zone}\\
\rho_v\, R_{\text{line}}\, s_v &\;\le\; B_z^{\text{eff}}/w_z \quad (\text{serial } v).
\label{eq:roof-lane}
\end{align}

\emph{Channels.} A cross-zone edge's demand on its channel is its traffic times its transfer size; the total over all edges through the channel must fit its bound.
\begin{equation}
\forall c,\ \sum_{(u,v)\ \text{on}\ c}\rho_v\, R_{\text{line}}\, t_{u,v} \;\le\; B_c .
\label{eq:roof-chan}
\end{equation}

\smallskip\noindent\textbf{M3: Capacity.} The data structures placed on a zone must fit within its total memory capacity.
\begin{equation}
\forall z,\ \sum_{v:\mu(v)=z} m_v \;\le\; \mathit{Cap}_z .
\label{eq:cap}
\end{equation}
 
\subsection{Prediction using a simulator}\label{sec:simulator}
A design study is a cheap loop. The designer writes the platform graph (a Python description) once and expresses each candidate design as a program graph (a JSON file). A Python simulator performs the mapping, evaluates the three metrics, and prints a feasibility verdict along with the bottleneck.

The bottleneck then suggests how to revise the design\,---\,sift earlier to reduce $\rho$, move a task to a zone with more headroom, pull a working set below a $\kap$, or trade a handoff for co-location. A re-run takes seconds, so a designer can examine dozens of placements before building any implementation. Like other models of computation~\cite{ram, pram}, \zram aims to be accurate to within constant factors rather than exact. When a new hardware version ships and platform constants shift, only the platform graph needs to be modified.

%% file: sections/platform.tex
\section{The BlueField-3 Platform Graph}
\label{sec:platform}

\begin{figure}[t]
  \centering
  \includegraphics[width=0.80\columnwidth]{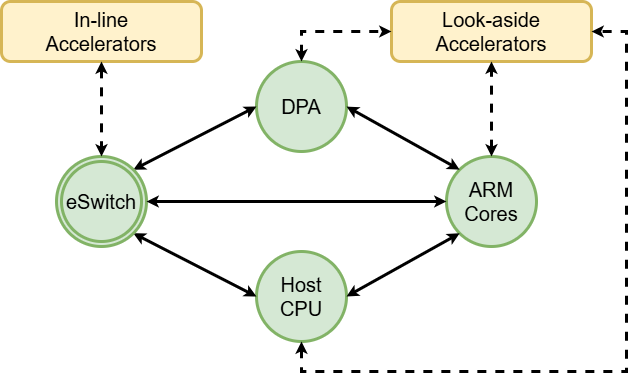}
  \vspace{-8pt}
  \caption{The BF-3 platform graph. Green zones (eSwitch, DPA, ARM cores, host CPU) and solid channels are the scope of this paper; yellow accelerators and dashed channels are shown to suggest additional zones \zram could expand to.}
  \label{fig:platform}
  \vspace{-12pt}
\end{figure}

We model the BF-3 with four primary zones (Fig.~\ref{fig:platform}). The eSwitch performs hardware matching at 400 Gbps, but has a very restrictive operation menu. The DPA sits directly on the datapath, but its cores are weak: even with its working set cached, its aggregate memory bandwidth runs roughly 8$\times$ below the host's. The ARM runs full Linux across 16 cores, but sits a hop off the fast path. The host is the strongest zone, but a full PCIe round trip away. No zone dominates.

Tables~\ref{tab:zones} and~\ref{tab:chan} report the BF-3's parameters. A few entries merit explanation. For the eSwitch$\leftrightarrow$ARM, eSwitch$\leftrightarrow$host, and ARM$\leftrightarrow$host channels, no public benchmarks exist. We bound each by the line rate. This is safe here: the underlying PCIe link exceeds it, and our case studies move only modest state per packet. We estimate the eSwitch's 8 GB memory capacity from driver limits: up to 8M flow rules, each with roughly 16 steering-table entries of 64 B. We measure the ARM and host memory bandwidths with LMbench streaming reads on our Intel Xeon Gold 6430 testbed.

\begin{table*}[t]\centering\small
\caption{BlueField-3 zone parameters. Flags: [e] measured on our testbed.}
\label{tab:zones}
\vspace{-8pt}
\begin{tabular}{@{}p{1.5cm}p{3.2cm}p{2.5cm}p{3.1cm}p{3.1cm}p{2.2cm}@{}}
\toprule
Zone & Operation menu & Cache threshold & Memory bandwidth & Memory capacity & Width \\
\midrule
eSwitch & match, meter, counter & --- & 400 Gbps~\cite{bluefield_3} & 8 GB [e] & 1~\cite{bluefield_3} \\
DPA & general compute & 1.5 MB~\cite{demystifying} & 8 Gbps / 200 Gbps~\cite{demystifying} & 1 GB~\cite{demystifying} & 256 (16$\times$16)~\cite{bluefield_3} \\
ARM & general compute & --- & 610 Gbps [e] & 32 GB~\cite{battle} & 16~\cite{battle} \\
Host & general compute & --- & 1{,}560 Gbps [e] & 256  GB~\cite{bluefield_3} & 16~\cite{bluefield_3} \\
\bottomrule
\end{tabular}
\vspace{-4pt}
\end{table*}

\begin{table}[t]\centering\small
\caption{BlueField-3 channel parameters. Flags: [b] bounded by
line rate; [e] measured on our testbed.}
\label{tab:chan}
\vspace{-8pt}
\begin{tabular}{@{}p{3.1cm}p{2.1cm}p{2.1cm}@{}}
\toprule
Channel & Bandwidth ($\rightarrow$) & Bandwidth ($\leftarrow$) \\
\midrule
Network $\leftrightarrow$ eSwitch & 400 Gbps~\cite{bluefield_3} & 400 Gbps~\cite{bluefield_3} \\
eSwitch $\leftrightarrow$ DPA & 44 Gbps [e] & 100 Gbps~\cite{demystifying} \\
eSwitch $\leftrightarrow$ ARM & 400 Gbps [b] & 400 Gbps [b] \\
eSwitch $\leftrightarrow$ Host & 400 Gbps [b] & 400 Gbps [b] \\
DPA $\leftrightarrow$ ARM & 200 Gbps~\cite{demystifying} & 200 Gbps~\cite{demystifying} \\
ARM $\leftrightarrow$ Host & 400 Gbps [b] & 400 Gbps [b] \\
\bottomrule
\end{tabular}
\vspace{-4pt}
\end{table}

\noindent\textbf{Cache thresholds:} The $\kap$ rule of \autoref{sec:model}
yields one threshold: 1.5\,MB for the DPA. We record none for the ARM or host: a large per-packet state workload could push either's demand
$\rho_v R_{\text{line}} s_v$ past line rate and would earn a $\kap$ exactly like
the DPA, but our case studies keep per-packet state small.

\noindent\textbf{Width:} Width splits a zone's two
speeds\,---\,parallel work sees the aggregate bound, a single packet's serial work sees one lane. The DPA's
aggregate divides across 256 hardware lanes (16$\times$16), the ARM's across 16. Thus per lane\,---\,and hence for serial code\,---\,the ARM is far faster, while on embarrassingly parallel work the DPA's aggregate bandwidth is competitive. The DPA wins on proximity and parallelism, the ARM on per-core strength and memory.

\noindent\textbf{Placement intuition:} Two features drive most placement decisions. First, the DPA is the only zone with \emph{two} bounds an order of magnitude apart; every DPA placement lives or dies by whether its working set fits within the 1.5 MB $\kap$. Second, measurements upend the intuition that on-die means wide: the eSwitch$\to$DPA ingress is the narrowest channel in Table~\ref{tab:chan}. DPA-resident buffers offer only 44 Gbps while ARM-memory buffers~\cite{demystifying}
provide line rate. Thus decomposing a program across zones is governed by
budgets and buffer placement, not proximity.

\noindent\textbf{Accelerators:}
The BF-3's fixed-function accelerators (inline crypto; look-aside RegEx,
compression, DMA) are zones. The model extends naturally\,---\,a vertex with a \emph{one-entry
menu} and channels from invoking zones. A program that
invokes one is doing exactly that function\,---\,there is no contested placement decision. Therefore, we set accelerators aside.

\noindent\textbf{All-eSwitch designs:} Table~\ref{tab:zones} shows that the eSwitch has 8 GB of
flow-table space, so why place work elsewhere? Unfortunately, the
eSwitch's menu is restrictive and cannot represent complex data structures and operations. Also, the eSwitch's flow-tables are heavily shared: they serve L2/L3 forwarding, tenant isolation, ACLs, and telemetry; filling them for one program impacts the rest. Placement is therefore dictated first by capability and second by stewardship.

%% file: sections/ddos.tex
\section{Case Study: Hierarchical DDoS Detector}
\label{sec:ddos}

\begin{figure}[t]
  \centering
  \includegraphics[width=0.9\columnwidth]{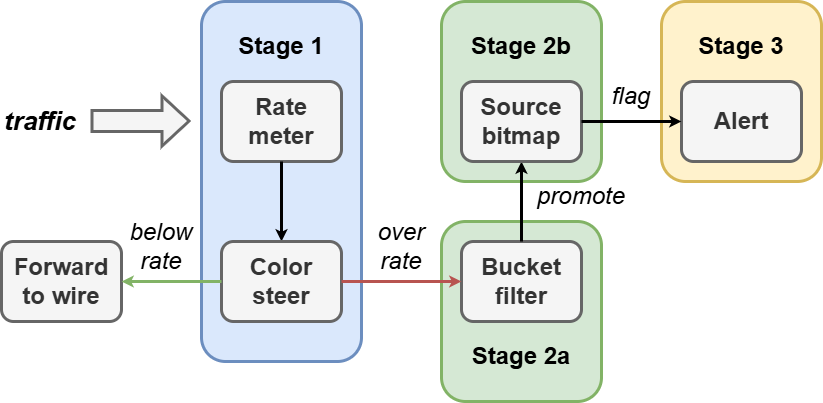}
  \vspace{-8pt}
  \caption{The DDoS detector program graph. Grey nodes are tasks. Tasks are grouped into stages for exposition.}
  \label{fig:ddos}
  \vspace{-8pt}
\end{figure}

DDoS detection is a vast area~\cite{ddos_survey, reiher}. Our simple algorithm
adapts ideas in \cite{netsift, estan}. We target volumetric attacks in which
many distinct sources converge on one destination (e.g., DNS reflection).
Volume alone is not a reliable signal---legitimate traffic can also be
heavy---so the detector measures \emph{source diversity toward one
destination}, and it is organized around sifting: cheap hardware checks run on
all traffic, and progressively more expensive software runs only on the small
residue that survives each check. The detector (Fig.~\ref{fig:ddos}) has three
stages plus an asynchronous host alert.

\noindent\textbf{Stage~1: meter and steer.} A hardware \emph{meter}
per protected destination---one exact-match rule each---marks a packet red iff
the destination's rate exceeds a threshold; a second pipe steers on the color:
green forwards out, red enters Stage~2.

\noindent\textbf{Stage 2a: bucket filter.} Each source is mapped into
one of eight buckets by its lower three bits. Per destination, the stage
keeps a deduplicated set of active buckets and their count; when the count
crosses a threshold the destination is \emph{promoted}, and a steering rule
sends its subsequent packets straight to Stage~2b. This per-packet
set-insertion must live in a \textsf{general compute} zone.

\noindent\textbf{Stage 2b: source bitmap.} Per promoted destination,
one packed 32-bit word: a 28-bit bitmap plus a sampling exponent $sr$. Each
source hashes to a bit, so the bitmap counts \emph{distinct} sources. When it
half-fills, the count is folded into a floor, the bitmap clears, and $sr$
increments---halving the sampling rate so the word never saturates; the
estimate $\max(\text{floor},\,\text{popcount}\cdot 2^{sr})$ is monotone.
Crossing a second threshold flags the destination.

\noindent\textbf{Stage 3: host alert.} On a flag, (dst, estimate,
timestamp) is shipped to the host and logged; in production the host installs
eSwitch drop or rate-limit rules to divert attack traffic.

\smallskip We modeled this program graph under three placements of Stages 2a/2b (Table~\ref{tab:placements}) and ran the simulator against the BF-3 platform graph on a real DNS-DDoS trace of 30,556,090 packets and 5,074,414 flows.


\smallskip\noindent\textbf{Verdict: feasible.} M1 places the stages: metering
and steering are native eSwitch actions, but the per-packet set-insertion of
Stages 2a/2b is not in the eSwitch menu, so both move to the DPA (all-eSwitch
fails M1 outright). M3 clears easily---151 meters and two steering pipes fit
the eSwitch's $\approx$8\,GB budget, and the bucket counters and distinct-source bitmaps exactly fit 1.5 MB of the
1 GB DPA region. The \emph{binding constraint is the DPA hot-set at 100\% of
$\kap$}.

\noindent\textbf{Comparing placements.} Each row of
Table~\ref{tab:placements} is one placement, scored by \zram; $\rmax$ is the
largest escalation fraction it can absorb before its first constraint
saturates, and the M2 cell names the placement's \emph{tightest} roofline term.
Stage~1 is pinned to the eSwitch by
capability, so the free variable is where Stages~2a/2b land. In the chosen
DPA/DPA row that tightest term is the working set against $\kap$
(1.5/1.5\,MB): feasible at the cached bound, but with zero cache margin.
Moving 2b (or both stages) to the ARM relieves the cache---the tables sit
comfortably in 32\,GB of DDR---at the price of a cross-zone hop per escalated
packet; the tightest remaining term is then Stage~1's own work, identical in
every row: roughly 56\,B of header and meter state touched per packet, i.e.,
267\,Gbps against the eSwitch's 400\,Gbps line-rate bound.

\noindent\textbf{Attack headroom.} The
maximum escalation fraction the DPA can absorb follows directly from the zone
roofline, Eq.~\eqref{eq:roof-zone}. Each escalated packet touches
$p\approx1{,}344$ bits (168\,B) of DPA state, derived from the trace, so
\begin{equation}
\rho\cdot 595\,\mathrm{Mpps}\cdot 1{,}344\,\mathrm{bits}\approx
\rho\cdot 800\,\mathrm{Gbps}.
\label{eq:demand}
\end{equation}
Resolving Eq.~\eqref{eq:resolve} against the cached versus uncached DPA
bounds,
\begin{equation}
\rmax^{\text{cached}}=\frac{200}{800}=0.25,\qquad
\rmax^{\text{uncached}}=\frac{8}{800}=0.01 .
\label{eq:rmax}
\end{equation}

\begin{table*}[t]
  \centering
  \small
  \setlength{\tabcolsep}{6pt}
  \renewcommand{\arraystretch}{1.3}
  \caption{Feasibility and attack headroom for three placements of the DDoS detector, computed by \zram from the platform and program graphs. All three pass M1--M3; the binding constraint and maximum escalation fraction $\rho_{\max}$ differ by placement.}
  \vspace{-8pt}
  \begin{tabular}{@{}cccc c l c c@{}}
    \toprule
    \textbf{Stage 1} & \textbf{Stage 2a} & \textbf{Stage 2b} & \textbf{Stage 3} &
      \textbf{M1 (capability)} & \textbf{M2 (roofline)} & \textbf{M3 (capacity)} &
      \textbf{$\bm{\rho_{\max}}$ (attack)} \\
    \midrule
    eSwitch & DPA & DPA & Host & \cmark &
      \cmark~: DPA $\kappa$: 1.5/1.5\,MB &
      \cmark &
      0.25 \\
    eSwitch & DPA & ARM & Host & \cmark &
      \cmark~: eSwitch: 267/400 Gbps &
      \cmark &
      0.50 \\
    eSwitch & ARM & ARM & Host & \cmark &
      \cmark~: eSwitch: 267/400 Gbps &
      \cmark &
      0.50 \\
    \bottomrule
  \end{tabular}
  \label{tab:placements}
  \vspace{-4pt}
\end{table*}

%% file: sections/cases.tex
\section{Additional Case Studies}
\label{sec:cases}

We briefly explore how \zram can be applied to decision-tree inference and RDMA traversal.

\subsection{Decision-tree inference}
Consider classifying traffic with a random forest: parse header features,
evaluate each tree's top levels, and descend into the deeper levels only for
packets the top cannot resolve~\cite{iisy, planter}. The design question is where the forest lives, and M2 answers it through the cache threshold. A forest's shallow levels are small\,---\,a handful of nodes per tree, kilobytes in total\,---\,so they sit comfortably under the DPA's 1.5 MB $\kap$. The full trees run to megabytes, and placed whole on the DPA they spill past $\kap$ and drop to the uncached rate, a $25\times$ penalty.

The state must therefore \emph{bifurcate}: the heavily used shallow prefixes belong on the DPA within its cache bound, and the rarely used deep tails live in ARM memory. The channel now binds. Every packet the DPA cannot resolve crosses DPA$\to$ARM and spends its transfer size against that budget, so the design lives or dies by the crossing fraction, which the required classification accuracy fixes. This is the sift-early logic of the DDoS detector, applied to a channel rather than a cache. \zram also makes a limit explicit: one might hope to shrink $\rho$ ahead of the DPA with a SYN-only eSwitch filter, but the eSwitch matches only on shallow header fields, not on the content an effective discriminant needs, so the sift cannot move earlier.

\subsection{Offloaded RDMA traversal}
A single one-sided RDMA operation offers little placement freedom, since the host must issue it and reap its completion regardless.
Freedom appears once a \emph{chain} of dependent operations is offloaded.
Consider walking a linked list in far memory~\cite{legoos} with one-sided RDMA, where each read returns the address of the next node.
Driven from the host, each hop is a separate command the host can issue only after the previous read returns, so an $n$-node walk serializes $n$ round trips.
Offloaded, the host issues one command, the NIC follows the whole chain on its own, and a single completion returns the result\,---\,paying the host round trip once and amortizing it over the list.
The same shape underlies SmartNIC systems that originate messages or RPCs, such as iPipe~\cite{ipipe} and Xenic~\cite{xenic}.

\zram guides where the chain should run.
The traversal is serial\,---\,each hop waits on the previous pointer\,---\,so by P3 the per-lane rate governs, not aggregate throughput.
An ARM or host core walks the chain far faster per hop than a slow DPA thread, favoring the ARM for a latency-bound chase; the DPA wins only when many independent chains run in parallel.
Where the data lands is a separate decision: the list lives in far memory, so the driving zone need not hold it, and the NIC's DMA steers each fetched node into whatever zone has capacity while a small zone issues the operations.
On the DPA, the cache threshold sets the rate\,---\,if the nodes the traversal revisits fit under $\kap$ it runs cached, otherwise it drops to the uncached bound.

%% file: sections/patterns.tex
\section{Design Patterns}
\label{sec:patterns}
\zram does more than score a finished placement; it also guides how to improve one.
When the simulator exposes a bottleneck (\autoref{sec:simulator}), the fix is usually one of a few recurring moves; re-running the three metrics confirms whether the move helped.
We distill seven such moves that recur across our case studies and in prior SmartNIC and switch designs.
Each activates a specific lever in the model: P1, P6, and P7 shrink or redirect the traffic fraction $\rho$; P4 and P5 hold a working set under the cache threshold $\kap$; P2 resolves a capability failure; P3 matches a task to a zone's width.


\noindent\textbf{P1: Sift early.} Do cheap, common-case checks close to the wire so only a small residue continues to later stages~\cite{netsift}, shrinking the traffic fraction $\rho$ at every hop. In the DDoS detector, 99.99\% of packets never leave the eSwitch.

\noindent\textbf{P2: Offload to a specialist.} When one operation dominates,
move it to the fixed-function zone built for it and keep everything else on a
general core~\cite{ipipe,floem}. Use this when a zone fails capability on a compute-intensive transform\,---\,crypto, regex, hashing\,---\,that an accelerator runs at wire speed.

\noindent\textbf{P3: Match width to work.} Parallel packet streams are well matched to the
DPA's hundreds of slow lanes~\cite{demystifying}; serial control loops require a few fast ARM or host
cores. Inspect the aggregate bound for parallel work and the per-lane floor for
serial work to determine which zone is a better fit.

\noindent\textbf{P4: Fit under threshold.} Keep a zone's working set below its cache threshold $\kap$ to reap the fast bound~\cite{cram}; spill past it and throughput collapses ($25\times$ on the DPA). 

\noindent\textbf{P5: Bifurcate state.} Split state into a small highly used part that
remains within the $\kap$ bound, and a less used portion in zones with large DRAM.

\noindent\textbf{P6: Steer by a shallow discriminant.} Let the eSwitch classify
on a few header bits~\cite{rmt}---a color, a method ID, a prefix---and route each class
straight to the zone that handles it, 
thereby using hardware matching for software dispatch.


\noindent\textbf{P7: Escalate, then bypass.}  Change placement when conditions change. For example, on a DDoS alert, the host may install eSwitch drop rules to divert attack traffic~\cite{ddos_survey}. 



%% file: sections/related.tex
\section{Related Work}
\label{sec:related}

\noindent\textbf{Hardware models}: \cram~\cite{cram} is a 
model for \emph{single-pipeline} RMT~\cite{rmt} router chips. 
\cram is a special case of \zram using one line-rate lookup zone with $\rho=1$.  \zram's contribution is
the graph structure \cram lacks along with heterogeneous zones.
LogP~\cite{logp} and BSP~\cite{valiant} model homogeneous hardware.

\noindent\textbf{Offload systems:} iPipe~\cite{ipipe}, Xenic~\cite{xenic}, Floem~\cite{floem}, Azure AccelNet~\cite{accelnet}, LineFS~\cite{linefs}, and FaRM~\cite{farm} each pick a placement empirically, per application;
\zram predicts feasibility \emph{before} implementation. Alkali~\cite{alkali} ports programs across NICs, operating \emph{below} \zram at the instruction level.


\noindent\textbf{Applications:} IIsy~\cite{iisy} and Planter~\cite{planter} map trees to match-action tables. White-Boxing RDMA~\cite{whitebox}, BluesMPI~\cite{bluesmpi}, and Palladium~\cite{palladium} inform the RDMA traversal study. Sifting was used for worm detection~\cite{netsift} but was implemented in a single software zone, not on many SmartNIC zones.

%% file: sections/open.tex
\section{Limitations}
\label{sec:open}

\noindent\textbf{Latency:} \zram scores bandwidth, not latency, by deliberate choice. Latency depends on queueing, invocation overheads, and handoff scheduling, and far more on the workload (arrival process, batch sizes, cache state) than a steady-state bandwidth bound does~\cite{dean}. Bandwidth, however, maps directly to the hardware a design needs, and hence its cost.

\noindent\textbf{Validation:} The DDoS detector experiment is a \emph{usability} result, showing the model is cheap and clear enough to apply in an afternoon. Confirming its predictions on real BF-3 hardware\,---\,that the named bottleneck saturates and throughput lands within constant factors\,---\,is the natural next step.

\noindent\textbf{Scope:} \zram currently assumes line-rate, steady-state designs on a single BF-3 running a single program, with feed-forward dataflow and per-zone own-memory. Making memory regions first-class vertices would disaggregate memory from processing~\cite{drmt}, and credit-based flow control or Receiver Not Ready (RNR) retry would need a cyclic formulation for their back-edges.

\noindent\textbf{Shared resources:} A \zram program is analyzed in isolation, but zones are shared. The DPA's $\kap$ is one global cache, and channels and memory controllers serve every flow and tenant, so a placement feasible by itself can fail once instances co-locate~\cite{fairnic}. Extending \zram means summing each zone's budget over co-resident programs, making feasibility a property of the entire workload.

%% file: sections/conclusion.tex
\section{Research Agenda}
\label{sec:conclusion}

\zram turns weeks of firmware-specific measurement into a script that runs in seconds and names the bottleneck resource. Across DDoS detection, decision trees, and RDMA, one insight recurs\,---\,\emph{a little sifting goes a long way}. Cutting $\rho$ upstream, at the eSwitch, relaxes every downstream roofline at once. And because a new device is just a new platform file, the same
approach extends to Intel IPU E2200~\cite{ipu_e2200}, AMD Pensando Salina 400~\cite{pensando_salina_400}, and beyond.

Beyond improving the model itself, \zram suggests a broader research agenda.

\noindent{\bf Validate on hardware:} Our immediate next step is to implement the
DDoS detector's placements on a real BlueField-3 and test whether the
bottleneck \zram names is the one that saturates first, and whether measured
throughput and $\rmax$ land within constant factors of the predictions.

\noindent{\bf Refine with more applications:} Other offloads (e.g., RPCs, TCP, transaction processing) will help evolve and refine \zram.

\noindent{\bf Automate design:} A compiler could pair \zram's cost model with an abstract functional spec (e.g., a DSL) and emit an implementation plan on the platform graph using the design idioms and integer linear programming~\cite{jose}.


\noindent{\bf Inform SmartNIC design:} 
Revealing which resources bottleneck a design, and why, can prioritize hardware upgrades. If the DPA binds, \zram indicates whether to strengthen its cores or its memory hierarchy (e.g., L2 cache).